\documentstyle[12pt,epsfig]{article} \author{Giovanni Punzi\\ INFN-Pisa}
\title{Notes on statistical separation of classes of events}
\date{January 23, 2003}
\begin{document} \maketitle

\section*{Introduction}

A common problem is that of separating different
classes of events in a given sample. One may want to separate some
"signal" from one or more "background" sources, or simply distinguish 
between different classes of signal events. There are several instances where
one cannot or does not wish to separate by means of cutting, and
instead wants to do a {\em statistical separation}. This means to be
able to calculate the number of events in each category that are
present in the given sample, and maybe measure some other
characteristics of each class, without explicitly labeling each
individual event as belonging to a particular category. For this to
be possible, one needs some observables that have different
distributions for each class of events. 

The purpose of this note is to define some criteria for
quantifying the resolution achievable in statistical separation,
given the distributions of the observables used to this purpose. 
One can use this to:

\begin{itemize}
\item quote the separation power of an observable in a compact way

\item quickly evaluate the expected resolution on extracting the
fractions of events in each category before actually performing any
fit

\item decide the optimal variables to use in separation when there
are several choices

\end{itemize}

\section*{Separating contributions}

Suppose your sample contains $n$ different classes of events, each
contributing a fraction $f_i$ of the total, and let $x$ be some
observable (which may be multidimensional) that is supposed to
distinguish between those events. The probability distribution of $x$
for our sample will be:

\begin{equation}
	p_{tot}(x|f) = \sum_{i=1,n}{f_i p_i(x)}
\end{equation}

where $p_i(x)$ is the pdf of $x$ for events of type $i$, and it is
assumed here to be perfectly known (any uncertainty in the $p_i(x)$
would contribute a systematic uncertainty to the final
results
). 

The most basic informations one wishes to extract from the sample of
data at hand is the values of the fractions $f_i$; we can therefore
take the resolution in extracting the $f_i$'s as the measure of the
separating power of the observable $x$.

The sum of all $f_i$ must be 1 in order for the overall distribution
to be correctly normalized, so there are actually only $n-1$ free
parameters to be evaluated; let's put arbitrarily
$f_n=1-\sum_{i=1,n-1}{f_i}$.

The resolution in estimating the $f_i$'s can in principle be measured
by setting up a Maximum Likelihood fit procedure, and repeating it on
a sufficient number of MonteCarlo samples to evaluate the spread of
results around the input values. You can also look at the resolutions
returned by your favorite fitter program, but it is important to
remember that those numbers are only approximate estimates of the
actual resolution achieved, especially when statistics is low and/or
the likelihood function is less than regular, so it is useful to be
able to calculate them indipendently. This is also a good
cross--check that the fit is actually doing what you want and that
its error estimates are sound.

A standard way to evaluate the resolution expected from a measurement
before actually carrying it out is to look at the Minimum Variance
Bound\cite{Eadie}:

\begin{equation}
cov(\hat{\mu_i}, \hat{\mu_j}) = - \left[ E
\left[\frac{\partial^2{\log {\cal L}}}{\partial{\mu_i}
\partial{\mu_j}} \right] \right] _{ij}^{-1}
\end{equation}

 this is an upper bound to the precision that can be achieved,
whatever the estimation procedure used. Whenever the problem is
sufficiently regular, the ML estimator gets in fact very close to
this limit. 

Luckily enough, the MVB for our problem can be written down in a
pretty simple form: the covariance matrix of the $n-1$ independent
$f_i$ parameter estimates is:

\begin{equation} cov(f_i,f_j) = \frac{1}{N} \left[\int 
    \frac{\left( p_i(x) - p_n(x) \right) \,\left( p_j(x) - p_n(x)
\right) }
      { p_{tot}(x|f)}\,dx \right] _{ij}^{-1}
\label{eq:cov}
\end{equation}    

(remember that the fraction $f_n$ associated to distribution $p_n(x)$
is determined from the other $f_i$'s). Note that in this formula the
symbol $x$ may stand for a set of many variables, discrete and/or
continuous, and the integrals extend over the whole $x$ domain. 

For a 2--component sample, there is only one fraction $f=f_1$ to be
evaluated, and the result is particulary simple:

\begin{equation}
 \sigma^2(f) = \frac{1}{N} \left(\int \frac{{\left( p_1(x) - p_2(x)
\right) }^2}
     { f p_1(x) + (1-f) p_2(x)}\,dx \right)^{-1} 
\label{eq:2comp}
\end{equation}    

This is the quantity you want to minimize in order to achieve the
best possible statistical separation.

In the limiting case of the different classes of events being totally
separated in $x$, that is, the $p_i(x)$ having zero overlap, the
uncertainties on $f_i$ come just from the statistical fluctuations of
the distribution of the events amongst classes due to finite sample
size, and eq.~\ref{eq:2comp} becomes:

\begin{equation}\label{eq:Binom} \sigma_{best}^2(f) =
\frac{f(1-f)}{N} \end{equation} 

which is the familiar result from the Binomial distribution.

It is particularly convenient to use the ratio of the resolution
(\ref{eq:2comp}) to the limit resolution (\ref{eq:Binom}), in order
to quote the separation power of the observable $x$ as an
adimensional quantity:

\begin{equation}
 s = \sigma_{best}(f)/\sigma(f) = \sqrt{f (1-f) \int \frac{\left(
p_1(x) - p_2(x) \right)^2}{p_{tot}(x|f)}\,dx }
\end{equation}

This is indipendent from the sample size $N$, and tells you at a
glance the power of the $x$ observable in separating the samples,
from $0$ (no separation) to $1$ (absolute maximum achievable with the
given sample).
This quantity is more informative than common expressions like
"$n$-sigma separation" or "curves overlap by $xxx$\%", as it tells
you exactly how good the observable $x$ is in separating the events,
and it is valid whatever the shape and the dimensionality of the
distributions involved.

\section*{Examples}

A simple and common example is the separation between two
1-dimensional gaussian distributions of same sigma. The above
quantity $s$ is easily evaluated by numerical integration. Note that
$s$, as it generally happens for resolutions, depends on the true
value of the fractions $f_i$. Figure~1 shows $s$ as a function of the
distance, in units of sigma, between the mean values of the two
gaussians, and the different curves are for different values of $f$.
From this graph you can read, for instance, that a separation of 1
sigma between roughly equally populated samples gives you a
resolution on the relative fractions slightly more than a factor of
two ($1/0.45$) worse than ideal, that is to say, the sample is
statistically equivalent to a fully separated sample of size smaller
by a factor $0.2=0.45^2$.

\begin{figure}[h]
\begin{center}
\fbox{ \includegraphics[width=\textwidth]{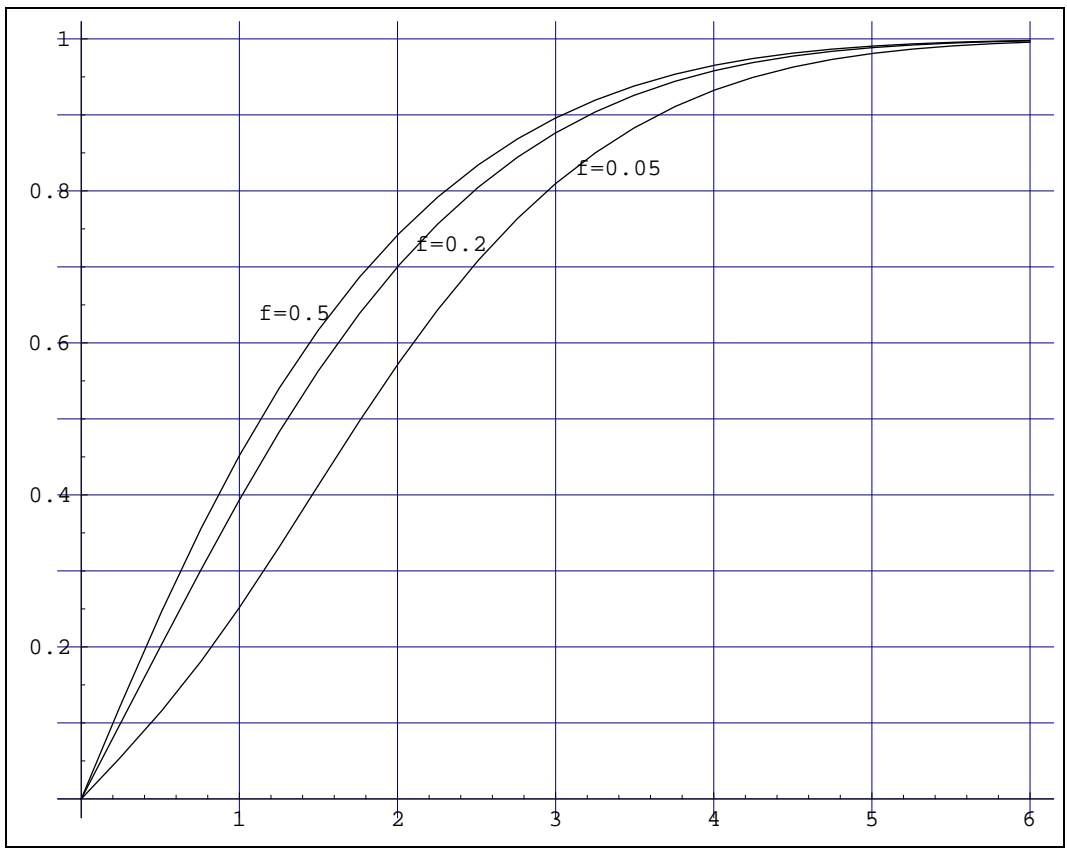}}
\caption{\label{fig:gsep} Separation power between two gaussians, as
a function of their distance}
\end{center}
\end{figure}

\sloppy

\end{document}